\documentclass[useAMS,usenatbib]{mn2e}

\usepackage{natbib}
\usepackage{graphicx}
\usepackage{amssymb}
\usepackage{xcolor}
\usepackage{caption}
\usepackage{lipsum,graphicx,multicol}
\usepackage{float}

\usepackage{subcaption}

\title[What powers galactic outflows: SB or AGN?]{What powers galactic outflows: nuclear starbursts or AGN?}
\author[ ]
{W. Ishibashi$^{1}$\thanks{E-mail: wako.ishibashi@physik.uzh.ch} and A. C. Fabian$^{2}$ 
\footnotemark[0]\\
\footnotemark[0]\\
$^{1}$Physik-Institut, Universitat Zurich, Winterthurerstrasse 190, 8057 Zurich, Switzerland \\
$^{2}$Institute of Astronomy, Madingley Road, Cambridge CB3 0HA 
}

\begin{document}

\pdfminorversion=4

\date{Accepted ? Received ?; in original form ? }

\pagerange{\pageref{firstpage}--\pageref{lastpage}} \pubyear{2012}

\maketitle

\label{firstpage}

\begin{abstract}
Galactic outflows can be powered either by nuclear starbursts (SB) or active galactic nuclei (AGN). It has been argued that extreme starbursts can power extreme outflows, without the need to invoke AGN feedback. However, contributions from past and/or hidden AGN activity cannot be ruled out. Here, we constrain the potential role of the central black hole in driving powerful outflows in starburst galaxies (with no sign of ongoing AGN activity). We examine whether the galactic outflows can be explained by AGN luminosity evolution in the framework of our AGN `radiative dusty feedback' scenario. We show that the outflow energetics of starburst galaxies in the local Universe can be quantitatively reproduced by power-law and exponential luminosity decays, coupled with radiation trapping. Likewise, a combination of heavy obscuration and mild luminosity decay may account for the energetics of galactic outflows observed in dusty star-forming galaxies in the early Universe. We discuss different physical arguments for SB vs. AGN outflow-driving, and conclude that the latter can have a major impact on the evolution of galaxies. 
\end{abstract} 



\section{Introduction}
\label{Section_Introduction}

Galactic outflows play fundamental roles in regulating the evolution of galaxies over cosmic time. Powerful outflows on galactic scales are now commonly observed in both star-forming galaxies and active galactic nuclei (AGN) host galaxies at low and high redshifts. Their multi-phase nature is revealed by detections in the ionised, neutral, and molecular gas phases \citep[e.g. see the recent reviews by][]{Veilleux_et_2020_review, Laha_et_2021}. Molecular outflows are of particular importance, as the molecular gas is the primary fuel for star formation. Observations indicate that a significant fraction of the outflowing gas is in molecular form, with molecular outflows reaching high velocities on kpc-scales. The associated outflow energetics are characterised by large momentum fluxes and high kinetic powers \citep{Fiore_et_2017, Gonzalez-Alfonso_et_2017, Fluetsch_et_2019, Lutz_et_2020}. 

Despite the huge observational progress in quantifying the outflow dynamics and energetics, the physical nature of galactic outflows --and their driving mechanism-- remains elusive. In some cases, it is not even clear what the actual power source is: a nuclear starburst (SB) or a central AGN? In general, powerful AGN-driven outflows can reach high velocities ($v \gtrsim 1000$ km/s) on galactic scales, while SB-driven outflows typically have lower velocities ($v \lesssim 500$ km/s) and more modest energetics. Fast outflows with velocities exceeding $\sim$1000 km/s are difficult to explain in terms of SB-driven models based on pure stellar feedback \citep{Sharma_Nath_2013, Veilleux_et_2020_review}.  

However, it has been argued that extreme SBs can power extreme outflows, without the need to invoke AGN feedback \citep{Diamond-Stanic_et_2012, Sell_et_2014, Geach_et_2014, Geach_et_2018}. High-velocity molecular outflows have been observed in massive compact SB galaxies at intermediate redshifts of $z \sim 0.6-0.7$ \citep{Geach_et_2014, Geach_et_2018}. Since no significant AGN activity is currently detected at their centres, the observed outflows are attributed to the nuclear SB. These compact SB galaxies (with effective radius of $R_e \sim 100$ pc) are characterised by extreme star formation rate densities approaching the Eddington limit for the so-called maximal starburst ($\Sigma_\mathrm{SFR} \sim 10^3 \, \mathrm{M_{\odot} \, yr^{-1} \, kpc^{-2}}$) \citep[e.g.][and references therein]{Crocker_et_2018}. The observed modest outflow energetics could then be powered by stellar radiation pressure and may be consistent with driving by stellar feedback alone ---without requiring AGN energy input. On the other hand, contributions from heavily obscured AGN and/or past AGN activity cannot be ruled out. 

In the context of AGN-driven outflows, two physical models have been debated in the literature: wind feedback and radiation feedback. In the wind feedback scenario, quasi-relativistic winds originating from the inner accretion disc shock against the cold interstellar medium and drive large-scale galactic outflows \citep[][and references therein]{King_Pounds_2015}. The resulting outflow can be momentum-driven (on small scales) or energy-driven (on large scales) depending on whether the shocked gas is efficiently cooled or not. Large momentum boosts and high kinetic powers can be reached in the wind energy-driven regime, comparable to the values observed in molecular outflows. 

Another way of driving galactic outflows is via direct radiation pressure on dust \citep{Fabian_1999, Murray_et_2005, Thompson_et_2015}. The radiation-matter coupling is considerably enhanced in the presence of dust grains, leading to the development of large-scale outflows. We have previously shown how such AGN `radiative dusty feedback' may account for the broad range of energetics observed in galactic outflows, provided that radiation trapping is included \citep{Ishibashi_Fabian_2015, Ishibashi_et_2018a, Ishibashi_et_2021}. We have also discussed the effect of the AGN luminosity temporal evolution on the inferred energetics of galactic outflows \citep{Ishibashi_Fabian_2018}. Here, we constrain the potential AGN contribution in powering galactic outflows in star-forming galaxies without signs of current AGN activity. We examine whether the outflow energetics observed in such SB galaxies could be explained in terms of heavily obscured AGN and/or AGN luminosity decay models, by comparing with observational samples at both low and high redshifts. 

The paper is structured as follows. In Section \ref{Sect_RadDustyFeedback}, we briefly recall the basics of AGN radiative dusty feedback. We first consider the particular case of `fossil outflows' with extreme energetics that cannot be easily explained by neither SB-driving nor AGN-driving (Section \ref{Sect_Fossil_Outflows}). We next analyse the outflow energetics of local SB galaxies, which may be interpreted in terms of different AGN luminosity decay histories (Section \ref{Sect_Local_Galaxies}). Galactic outflows observed in dusty star-forming galaxies in the early Universe could also be powered by heavily obscured AGN coupled with mild luminosity evolution (Section \ref{Sect_Early_Galaxies}). We discuss the physical nature of the central power source ---SB or AGN--- in light of recent observational results (Section \ref{Sect_Discussion}) and conclude in Section \ref{Sect_Conclusion}. 


\section{AGN radiative dusty feedback}
\label{Sect_RadDustyFeedback}

AGN radiation pressure on dust can drive powerful outflows on galactic scales. Here, we briefly summarise the main features of our AGN radiative dusty feedback scenario [for a more complete description, see e.g. \citet{Ishibashi_Fabian_2015}]. 
We assume that radiation pressure sweeps up the surrounding dusty gas into an outflowing shell. The corresponding equation of motion is given by 
\begin{equation}
\frac{d}{dt} [M_\mathrm{sh} v] = \frac{L(t)}{c} (1 + \tau_\mathrm{IR} - e^{-\tau_\mathrm{UV}} ) - \frac{G M(r) M_\mathrm{sh}}{r^2} , 
\label{Eq_motion}
\end{equation}
where $L(t)$ is the central luminosity, $M_\mathrm{sh}$ is the shell mass, and $M(r) = \frac{2 \sigma^2 r}{G}$ is the total mass distribution assuming an isothermal potential with velocity dispersion $\sigma$.  
The infrared (IR) and ultraviolet (UV) optical depths are given by $\tau_\mathrm{IR,UV} = (\kappa_\mathrm{IR,UV} M_\mathrm{sh})/(4 \pi r^2)$, where $\kappa_\mathrm{IR}$=$5 \, \mathrm{cm^2 g^{-1} f_{dg, MW}}$ and $\kappa_\mathrm{UV}$=$10^3 \, \mathrm{cm^2 g^{-1} f_{dg, MW}}$ are the IR and UV opacities, with the dust-to-gas ratio normalised to the Milky Way value. A steady outflow can develop provided that the outward force due to radiation pressure exceeds the inward gravitational force. 

AGNs are known to be variable objects that can vary by several orders of magnitude in luminosity over a range of timescales \citep[e.g.][]{Hickox_et_2014}. If the accreting matter is disrupted and carried away by the radiation pressure-driven outflow, the central luminosity output may decline over time. 
We consider different forms of AGN luminosity decay, the exponential decay:
\begin{equation}
L(t) = L_0 e^{-t/t_d} \, , 
\label{Eq_L_EXP}
\end{equation}
where $L_0$ is the initial luminosity and $t_d$ is a characteristic decay timescale; and the power-law decay: 
\begin{equation}
L(t) = L_0 (1 + t/t_d)^{-\delta} \, , 
\label{Eq_L_PL}
\end{equation} 
where $\delta = 1$ is the power-law slope \citep[see also][]{Ishibashi_Fabian_2018}.
Physically, an abrupt fall-off and removal of accreting gas may lead to a rapid exponential luminosity decay; whereas a more gentle power-law luminosity decay may be expected when the accretion disc slowly dissipates on a viscous timescale. 
In the framework of the AGN wind feedback scenario, \citet{Zubovas_2018} argues that exponential luminosity decays are rare, while a power-law luminosity decay (with slope $\sim 1$) can preserve the correlations between outflow properties and AGN luminosity.    

The energetics of galactic outflows can be quantified by three parameters: the mass outflow rate ($\dot{M}$), the momentum flux ($\dot{p}$), and the kinetic power ($\dot{E}_k$). By analogy with the observational works, we define the three quantities in the so-called thin shell approximation as \citep[e.g.][]{Gonzalez-Alfonso_et_2017}
\begin{equation}
\dot{M} = \frac{M_\mathrm{sh}}{t_\mathrm{flow}} 
= \frac{M_\mathrm{sh} v}{r} , 
\label{Eq_Mdot}
\end{equation} 
\begin{equation}
\dot{p}  = \dot{M} v 
= \frac{M_\mathrm{sh} v^2}{r} , 
\label{Eq_pdot}
\end{equation} 
\begin{equation}
\dot{E}_k = \frac{1}{2} \dot{M} v^2 
= \frac{M_\mathrm{sh} v^3}{2 r} .
\label{Eq_Ekdot}
\end{equation}
Furthermore, two derived quantities are often used in quantifying the outflow energetics: the momentum ratio ($\zeta$) and the energy ratio ($\epsilon_k$), defined as
\begin{equation}
\zeta = \frac{\dot{p}}{L/c} = M_\mathrm{sh} \frac{v^2}{r} \frac{c}{L} , 
\label{momentum_ratio}
\end{equation} 
\begin{equation}
\epsilon_k = \frac{\dot{E}_k}{L}  = \frac{1}{2} M_\mathrm{sh} \frac{v^3}{r} \frac{1}{L} .
\label{energy_ratio}
\end{equation}
While the outflow properties are observed at a certain distance from the centre (on kpc-scales), the AGN luminosity is derived from the currently measured nuclear flux. We note that this can have major effects on the observationally inferred values of the outflow energetics, as discussed below. 


\section{Fossil outflows: relics of past AGN?}
\label{Sect_Fossil_Outflows}

\begin{figure*}
\begin{multicols}{2}
    \includegraphics[width=0.8\linewidth]{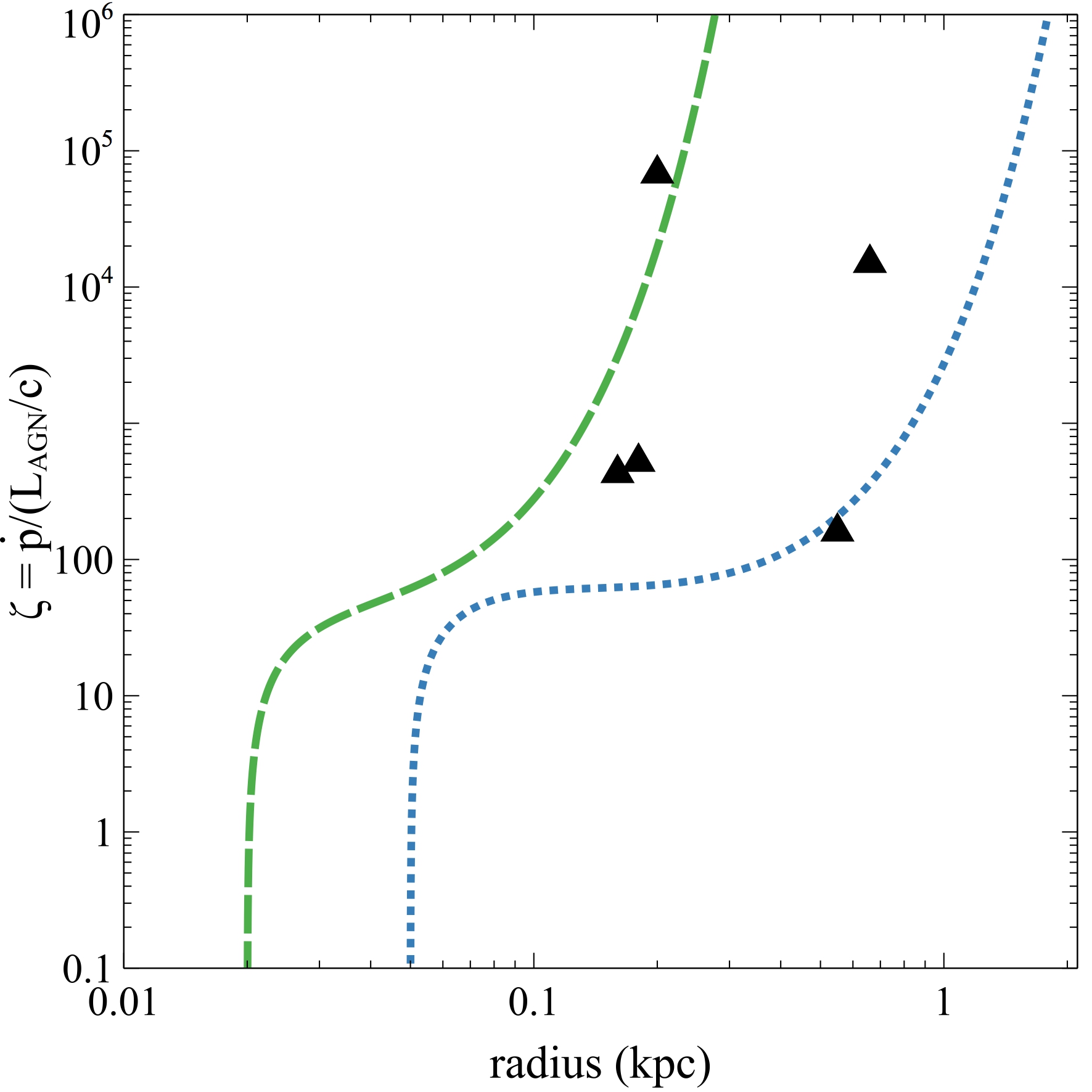}\par
    \includegraphics[width=0.8\linewidth]{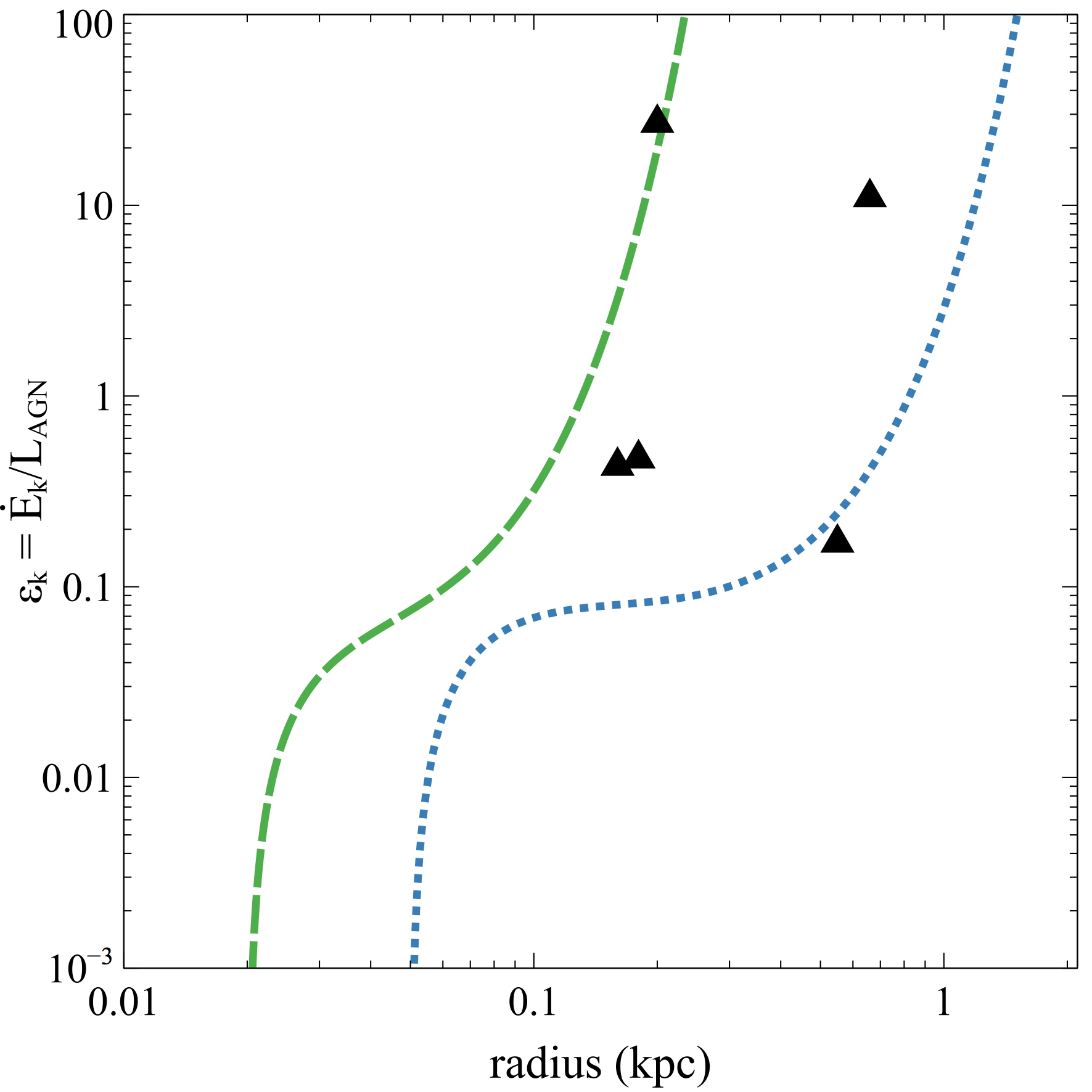}\par 
    \end{multicols}
\caption{ 
Momentum ratio (left-hand panel) and energy ratio (right-hand panel) of fossil outflows (black triangles) from the sample of \citet{Fluetsch_et_2019} compared to AGN-driven outflow models with exponential luminosity decay $L(t) = L_0 e^{-t/t_d}$: $L_0 = 1 \times 10^{46}$ erg/s with $t_d = 3 \times 10^4$ yr (green dashed), and $L_0 = 3 \times 10^{46}$ erg/s with $t_d = 2 \times 10^5$ yr (blue dotted). 
} 
\label{Fig_fossil_outflows}
\end{figure*}

Galactic molecular outflows with exceptionally high values of the momentum ratio ($\zeta > 100$) and energy ratio ($\epsilon_k > 0.1$) have been discovered in a number of local galaxies \citep{Fluetsch_et_2019}. Such extreme energetics cannot be accounted for by any plausible model of SB-driven or AGN-driven outflows (even AGN wind energy-driving is not sufficient). Instead, these may represent relics of past AGN activity that has since faded, and are prime candidates for `fossil outflows'. 

AGNs can undergo significant luminosity variations on short timescales, much shorter than the typical flow time of $t \sim$ Myr. Observationally, the outflow properties are measured at a certain radial distance from the centre (e.g. $r \sim 1$ kpc), while the central AGN luminosity is derived from the current flux observed in some waveband. 
These two quantities are then used to compute the outflow momentum ratio and energy ratio. However, it is unlikely that the central luminosity stays constant over the $\sim$Myr flow timescale, due to the variable nature of the AGN. If the AGN luminosity significantly decreases over this time span, the resulting momentum ratio and energy ratio can be largely overestimated \citep{Ishibashi_Fabian_2015}. 
Furthermore, the observed galactic outflow may not be identified as AGN-driven, as the central source has turned off. 
But there remains an alternative possibility: we currently observe a kpc-scale outflow, which was actually launched by an earlier AGN event in the past. 

In Fig. \ref{Fig_fossil_outflows}, we compare the fossil outflows observed in local galaxies to our AGN feedback models with exponential luminosity decay (equation \ref{Eq_L_EXP}). 
We see that the very high values of $\zeta > 100$ and $\epsilon_k > 0.1$ can be quantitatively accounted for by AGN radiative dusty feedback, due to the rapid exponential luminosity decay. Short decay timescales of $t_d \sim (10^4-10^5)$ yr are required to obtain the high energetics observed at small radii ($r \lesssim 1$ kpc) in fossil outflows. 
The associated AGN luminosities were likely much higher in the past, in some cases by a few orders of magnitude. 
This is a plausible assumption, given that accreting black holes are highly variable objects and AGN follow different luminosity evolution histories (see the discussion in Sect. \ref{Subsect_AGN_history}). Therefore fossil outflows could be relics of past AGN, and may be reproduced by radiative dusty feedback models with exponential luminosity decay.  


\section{Outflows in star-forming galaxies in the local Universe}
\label{Sect_Local_Galaxies} 

\begin{figure*}
\begin{multicols}{2}
    \includegraphics[width=0.8\linewidth]{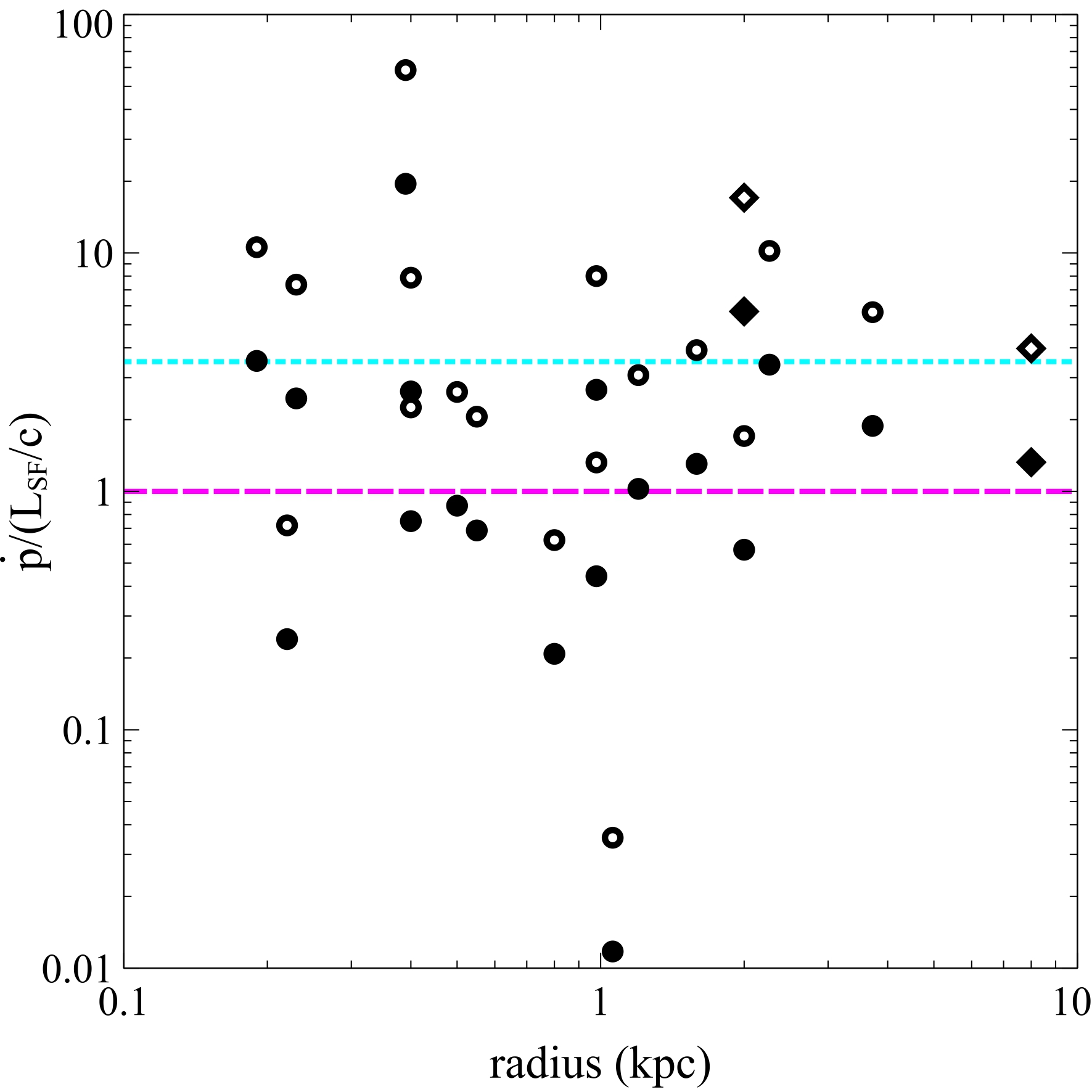}\par
    \includegraphics[width=0.8\linewidth]{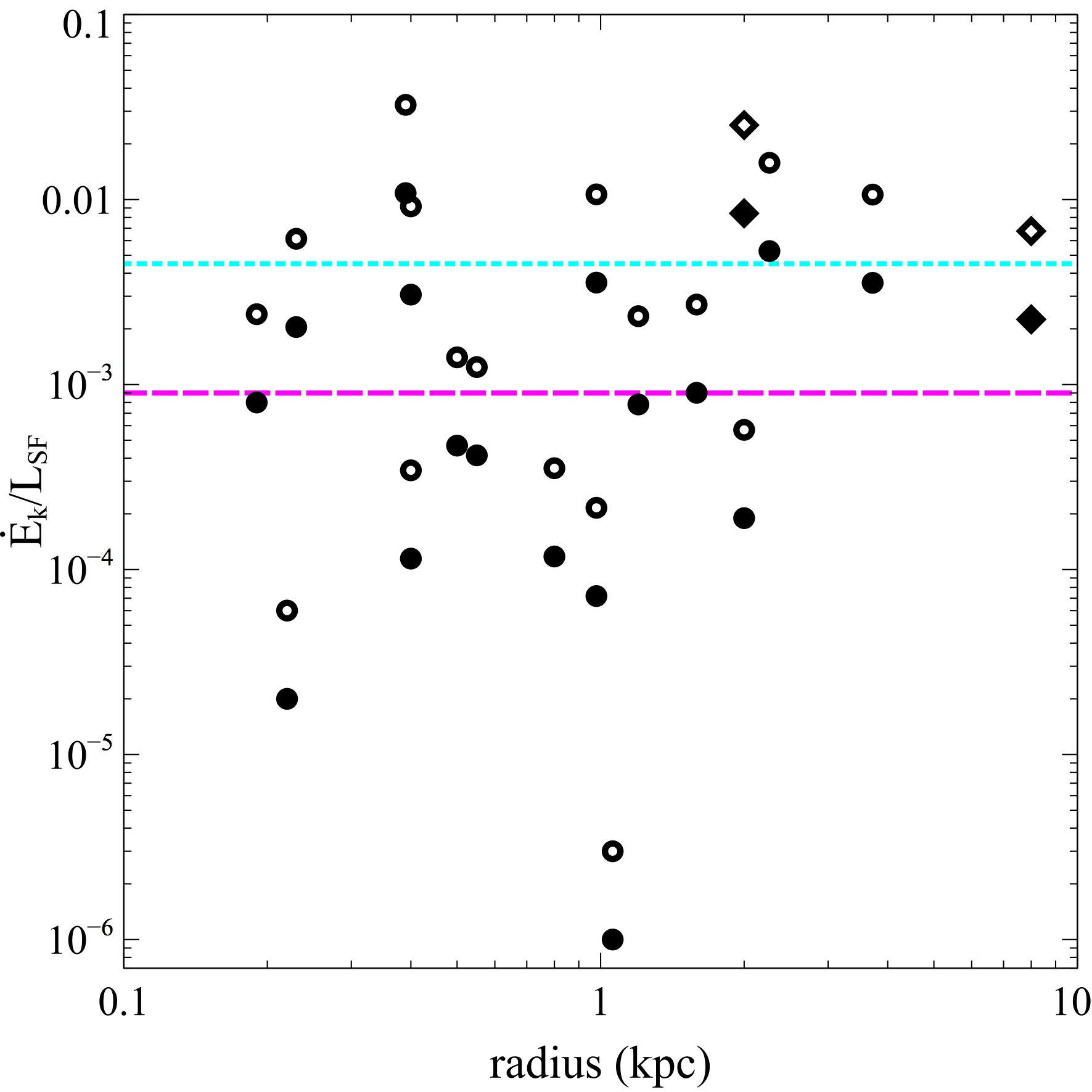}\par 
    \end{multicols}
\caption{
Outflow energetics of local star-forming galaxies (black dots) from the sample of \citet{Lutz_et_2020} and compact SBs at intermediate redshifts (black diamonds) from \citet{Geach_et_2014, Geach_et_2018}. 
Momentum ratio (left-hand panel) and energy ratio (right-hand panel) of molecular outflows (filled symbols) and total outflows (open symbols) with $\dot{M}_\mathrm{tot} \sim 3 \times \dot{M}_\mathrm{mol}$. Left-hand panel: the horizontal lines indicate the limits at $\dot{p} = L_\mathrm{SF}/c$ (magenta dashed) and $\dot{p} = 3.5 \, L_\mathrm{SF}/c$ (cyan dotted). Right-hand panel: the horizontal lines indicate a coupling fraction of $5 \%$ (magenta dashed) and $25 \%$ (cyan dotted), as described in the main text. 
} 
\label{Fig_local_SB_limits}
\end{figure*}

\begin{figure*}
\begin{multicols}{2}
    \includegraphics[width=0.8\linewidth]{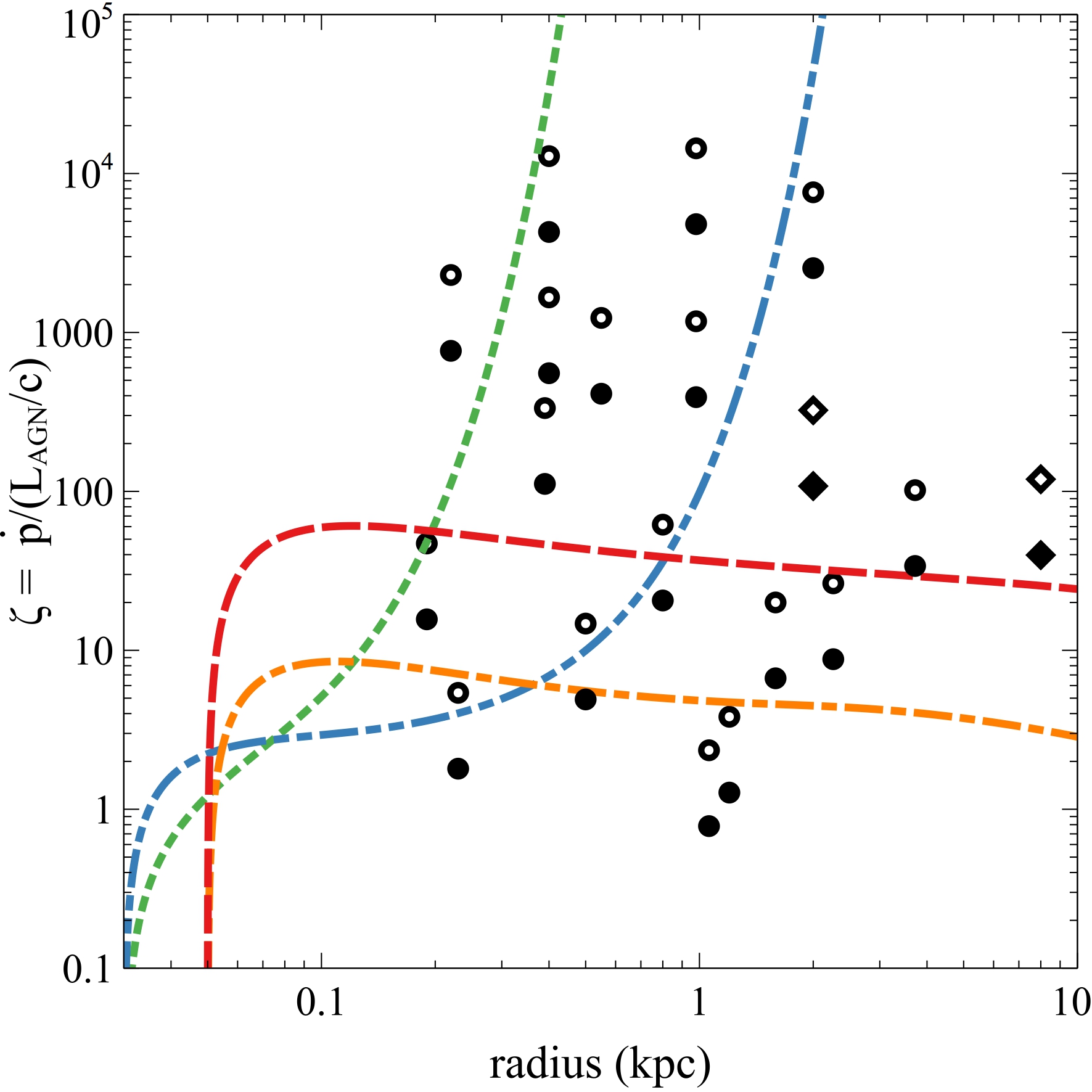}\par
    \includegraphics[width=0.8\linewidth]{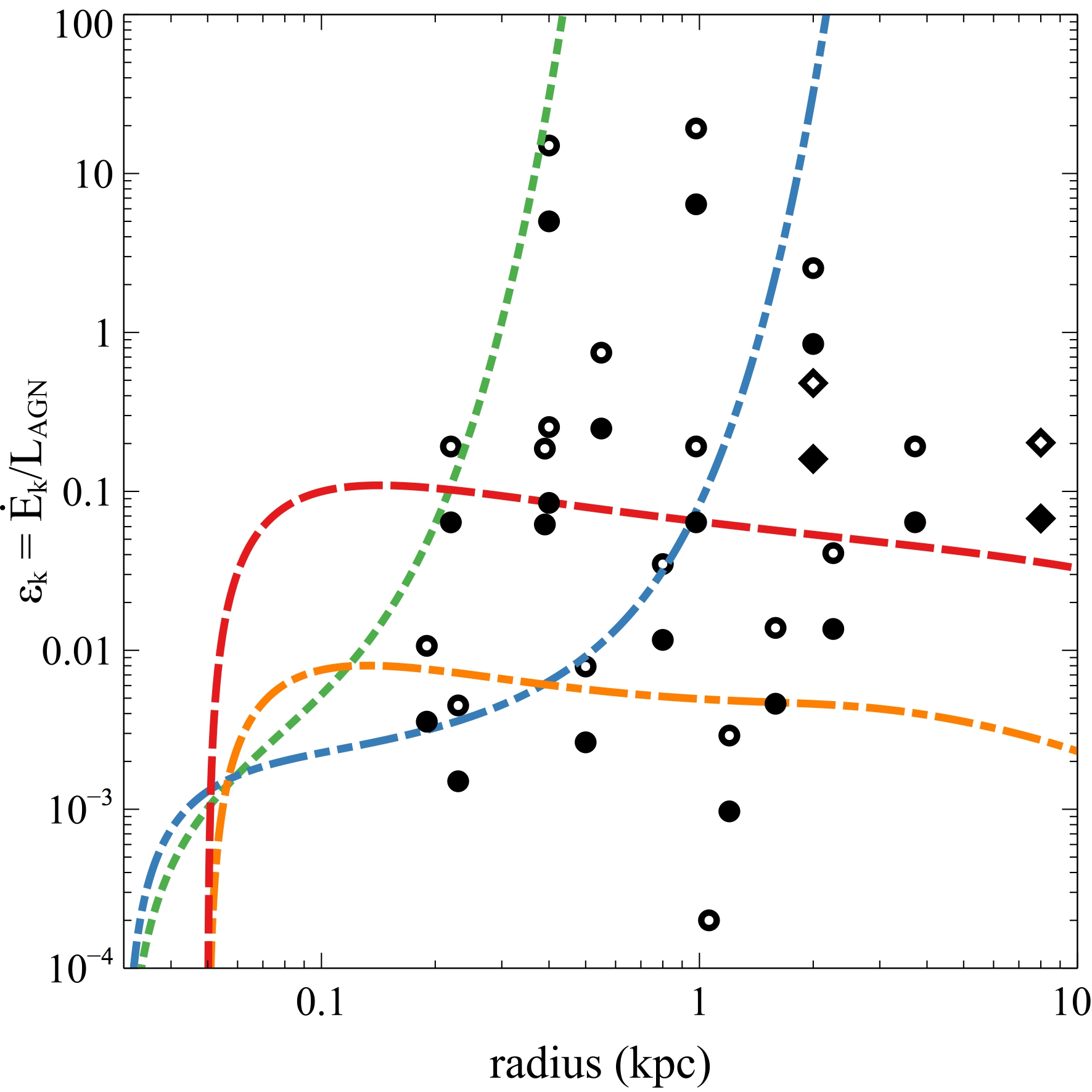}\par 
    \end{multicols}
\caption{ 
Outflow momentum ratio (left-hand panel) and energy ratio (right-hand panel) of SB galaxies in the local Universe (same symbols as in Fig. \ref{Fig_local_SB_limits}) compared to AGN radiative dusty feedback models with different luminosity decay forms.
Power-law luminosity decays $L(t) = L_0 (1 + t/t_d)^{-1}$: $L_0 = 1 \times 10^{46}$ erg/s with $t_d =  3 \times 10^5$ yr (orange dash-dot), and $L_0 = 6 \times 10^{46}$ erg/s with $t_d = 10^5$ yr (red dashed). 
Exponential luminosity decays $L(t) = L_0 e^{-t/t_d}$: $L_0 = 3 \times 10^{45}$ erg/s with $t_d = 3 \times 10^5$ yr (blue dash-dot-dot), and $L_0 = 3 \times 10^{45}$ erg/s with $t_d = 5 \times 10^4$ yr (green dotted). 
} 
\label{Fig_local_outflows}
\end{figure*}

We now consider the energetics of galactic outflows in local star-forming galaxies ($z < 0.2$) and compact starbursts at intermediate redshifts ($z \sim 0.7$). The local galaxies are classified as HII and LINER in \citet{Fluetsch_et_2019}, and the molecular outflow properties are based on \citet{Lutz_et_2020}. Fast molecular outflows with $v \sim 1000$ km/s on galactic scales are also reported in the two extremely compact SBs at $z \sim 0.7$ \citep{Geach_et_2014, Geach_et_2018}.

It has been suggested that stellar feedback can account for the outflow energetics observed in SB galaxies, without the need for AGN. 
In particular, \citet{Geach_et_2014} argue that the observed outflow momentum flux is consistent with the photon momentum output from the nuclear SB, and thus no AGN is required. In fact, the AGN contribution to the bolometric luminosity ($\alpha_\mathrm{AGN} = L_\mathrm{AGN}/L_\mathrm{bol}$) is estimated to be rather low (below 10 percent) in compact starbursts \citep{Sell_et_2014} and also very low ($\alpha_\mathrm{AGN} \ll 0.1$) in most local star-forming galaxies \citep{Fluetsch_et_2019}. 
However, AGN tracers based on X-ray or mid-IR luminosities are not always reliable, e.g. due to AGN short-term variability and/or heavy obscuration. Even if the current AGN contribution is negligible, one cannot rule out past AGN activity. 

Moreover, when computing the outflow energetics, one should also take into account the multi-phase nature of galactic outflows. 
In addition to the molecular component, other gas phases can contribute significantly to the total mass outflow rate in star-forming galaxies. Comparable mass outflow rates can be found in the neutral and ionised phases, such that the total mass outflow rate may be roughly three times higher, i.e. $\dot{M}_\mathrm{tot} \sim 3 \times \dot{M}_\mathrm{mol}$ \citep{Fluetsch_et_2019}. 
This will lead to a corresponding increase in the outflow energetics.    

Fig. \ref{Fig_local_SB_limits} shows the outflow momentum flux and kinetic power compared to the star formation output of SB galaxies in the local Universe. We observe that the momentum rate of molecular outflows is greater than the photon momentum flux due to stellar radiation pressure, i.e. $\dot{p} > (L_\mathrm{SF}/c)$, in a significant fraction of the sample (Fig. \ref{Fig_local_SB_limits}, left-hand panel). Including momentum deposition from stellar winds and supernovae, a continuous SB may provide a momentum boost of $\dot{p} \sim 3.5 \, L_\mathrm{SF}/c$ \citep{Veilleux_et_2005, Gonzalez-Alfonso_et_2017, Gowardhan_et_2018}. Several objects also exceed this limit, especially when considering the total mass outflow rates.  

Similarly, a significant number of sources have kinetic powers above the values expected from mechanical energy injection in SB galaxies (Fig. \ref{Fig_local_SB_limits}, right panel). The maximal mechanical luminosity provided by stellar winds and supernovae is about $\sim$$1.8 \%$ of the SB luminosity, of which only a fraction can be transferred to the bulk motion of the gas \citep{Veilleux_et_2005, Gonzalez-Alfonso_et_2017, Gowardhan_et_2018}. The coupling fraction may be between $\sim$$5\%$ and $\sim$$25\%$ depending on the different supernova environment configurations \citep{Walch_Naab_2015}. Several local SB galaxies seem to lie above these limits.  

We note that the outflowing gas is conservatively identified with the line wings only (rather than the entire broad component) in the sample of \cite{Lutz_et_2020}. In addition, if the AGN contribution to the bolometric luminosity is somewhat under-estimated (see the discussion in Sect. \ref{Subsect_IR_view}), the intrinsic starburst luminosity ($L_\mathrm{SF}$) could actually be lower. All these arguments suggest that pure SB-driving may not be enough to account for the observed outflow energetics, and that additional power sources are required in many star-forming galaxies in the local Universe. 

We therefore consider the possibility of AGN temporal evolution, with power-law and exponential luminosity decays. 
In Fig. \ref{Fig_local_outflows}, we show the resulting outflow momentum ratio and energy ratio compared to the observational measurements of local SB galaxies. The broad range of the observationally inferred outflow energetics can be quantitatively reproduced by AGN radiative dusty feedback with either power-law or exponential luminosity decays (with relatively short decay timescales $t_d \sim 10^4 - 10^5$ yr). Higher values of the momentum ratios and energy ratios may be obtained with exponential luminosity decays, while lower values of the outflow energetics can be accounted for by power-law luminosity decays.  

The physical conditions in the radiation pressure-driven outflows can range from IR-optically-thin to heavily IR-optically-thick cases (with initial IR optical depth covering the range $\tau_\mathrm{IR,0} \sim 0.1 - 110$). 
High column densities and large dust content can be found in the innermost regions of nuclear SBs and obscured AGNs. 
The most extreme cases are observed in the so-called `compact obscured nuclei' (CON), characterised by extreme column densities $N_\mathrm{H} \gtrsim 10^{25} \mathrm{cm^{-2}}$, which can be optically thick to IR and even sub-mm emission \citep[][and references therein]{Aalto_et_2019_review}. 

Observations indicate that large amounts of dust can be released in core-collapse supernovae, yielding high dust-to-gas ratios in supernova remnants \citep{Gomez_et_2012, Owen_Barlow_2015, Wesson_et_2015}. High values of the dust-to-gas ratio, of the order of $f_\mathrm{dg} \sim (1/50 - 1/20)$ have been inferred for sub-mm galaxies, dust-reddened quasars, and local ULIRGs \citep{Michalowski_et_2010, Banerji_et_2017, Gowardhan_et_2018}. 
Dust-to-gas ratios in the range $f_\mathrm{dg} = (1 - 5) \times f_\mathrm{dg,MW}$ are assumed in the radiative dusty feedback shell models (Fig. \ref{Fig_local_outflows}).

Here, we implicitly assume a constant dust-to-gas ratio throughout the outflow propagation, which is an optimistic assumption.
In fact, different dust destruction processes operate in galaxies, e.g. dust grains can be destroyed by sputtering and in shocks.
Radiation hydrodynamic (RHD) simulations of radiation pressure-driven shells suggest a rapid dust destruction in the AGN shock-heated outflow \citep{Barnes_et_2020}. Dust grains are mostly destroyed by thermal sputtering at high temperatures, such that the dust abundance could be significantly reduced at large galactocentric radii.

Radiation pressure feedback mainly operates on dense dusty gas, which is strongly radiating and remains cool ($T \lesssim 10^4$ K); while the outer gas swept up by the expanding shell may be temporarily shock-heated to high temperatures ($T \gtrsim 10^6$ K) and subsequently cool down. We note that the radiative cooling of dense gas should be very efficient. 
Once the shocked layer has cooled down, dust grains are also able to re-form via metal accretion \citep{Richings_et_2018}. 
Multi-dimensional RHD simulations indicate that surviving dusty clouds can be efficiently accelerated by radiation pressure \citep{Zhang_et_2018}.  
Dust grains embedded in outflowing clouds could even survive in hot winds --under certain favourable conditions-- as seen in some recent 3D hydrodynamic simulations \citep{Farber_Gronke_2022}. 

Dust is not only destroyed, but also created in galaxies. Core-collapse supernovae are known to release large amounts of dust, with a preference for dust grains of large sizes ($a_d \gtrsim 0.1 \mu m$) favouring a top-heavy grain size distribution  \citep[][and references therein]{Slavin_et_2020, Priestley_et_2022}. 
Larger dust grains are more likely to survive destruction by sputtering (as the sputtering timescale increases with grain size, $\tau_\mathrm{sput} \propto a_d$) compared to smaller grains that are easily destroyed in the supernova reverse shock. Thus a significant fraction of large dust grains can survive and enrich the surrounding medium. In addition, grain growth in the interstellar medium can also contribute to the dust mass increase at later times \citep{Michalowski_2015}, a process that can be further accelerated by turbulence \citep{Mattsson_2020}. 

Dust depletion could be partly compensated by dust replenishment, with the release of dust grains on large scales. 
We previously proposed the `AGN feedback-driven star formation' scenario, i.e. the possibility of star formation triggering within the AGN radiation pressure-driven outflows \citep{Ishibashi_Fabian_2012}. Such star formation occurring inside galactic outflows has now been observationally confirmed in a number of local galaxies \citep{Maiolino_et_2017, Gallagher_et_2019}. 
In this framework, new dust can be released at large radii, when massive stars (formed within the outflowing shell) explode as supernovae at the end of their lifetimes. 
Fresh dust can then be injected into the surrounding environment, sustaining the propagation of dusty outflows and further contributing to the overall AGN feedback process.


\section{Outflows in dusty star-forming galaxies in the early Universe}
\label{Sect_Early_Galaxies}

\begin{figure*}
\begin{multicols}{3}
    \includegraphics[width=\linewidth]{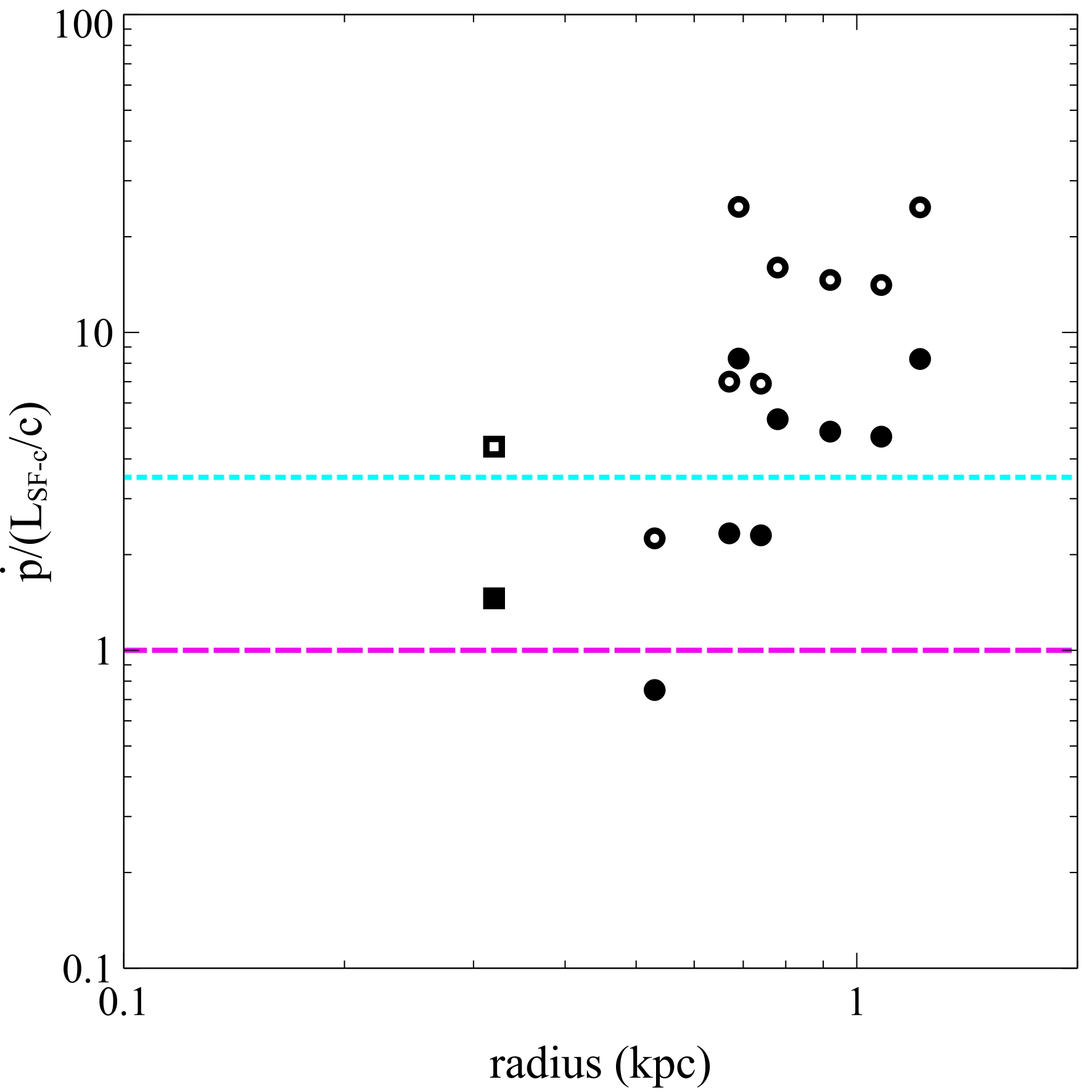}\par 
    \includegraphics[width=\linewidth]{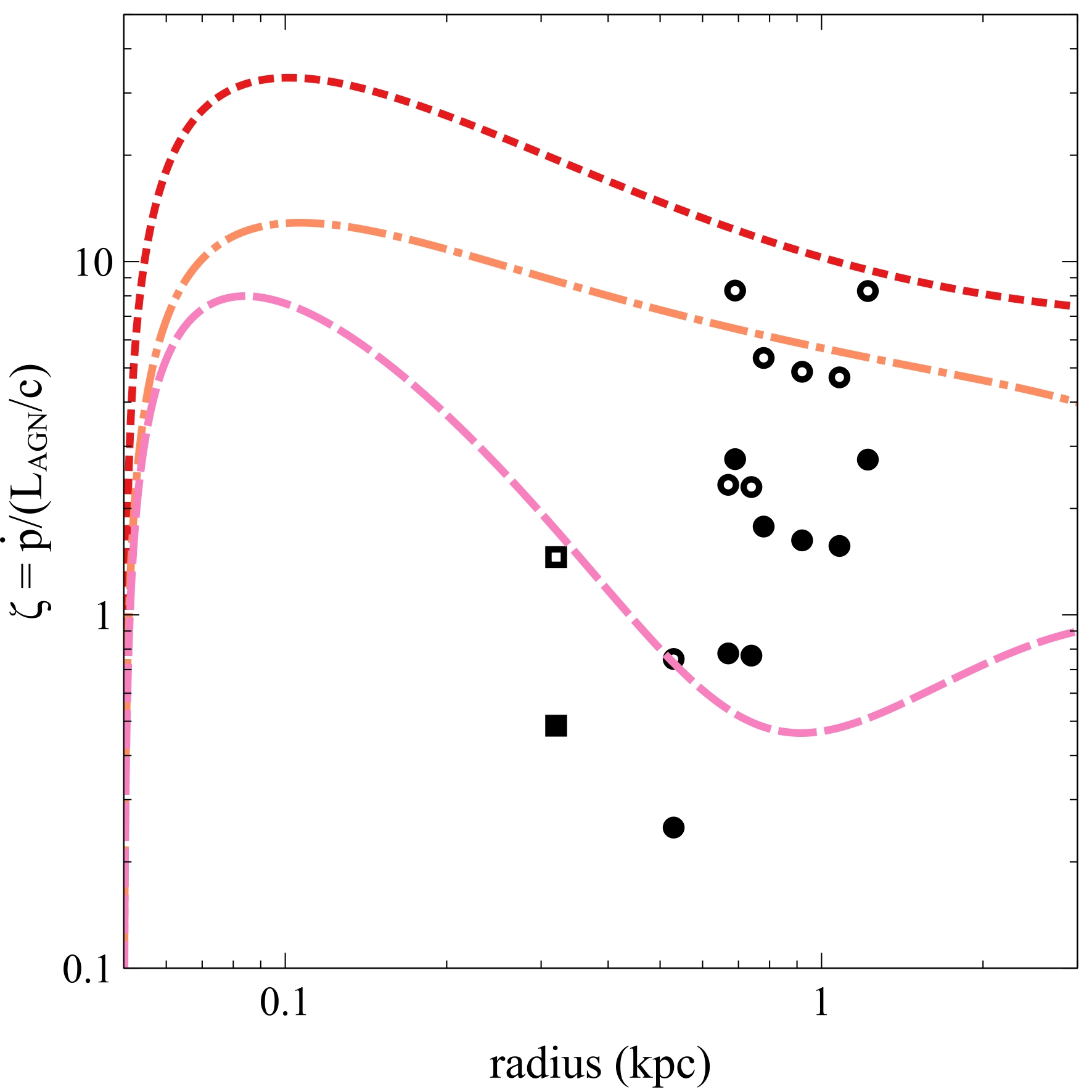}\par 
    \includegraphics[width=\linewidth]{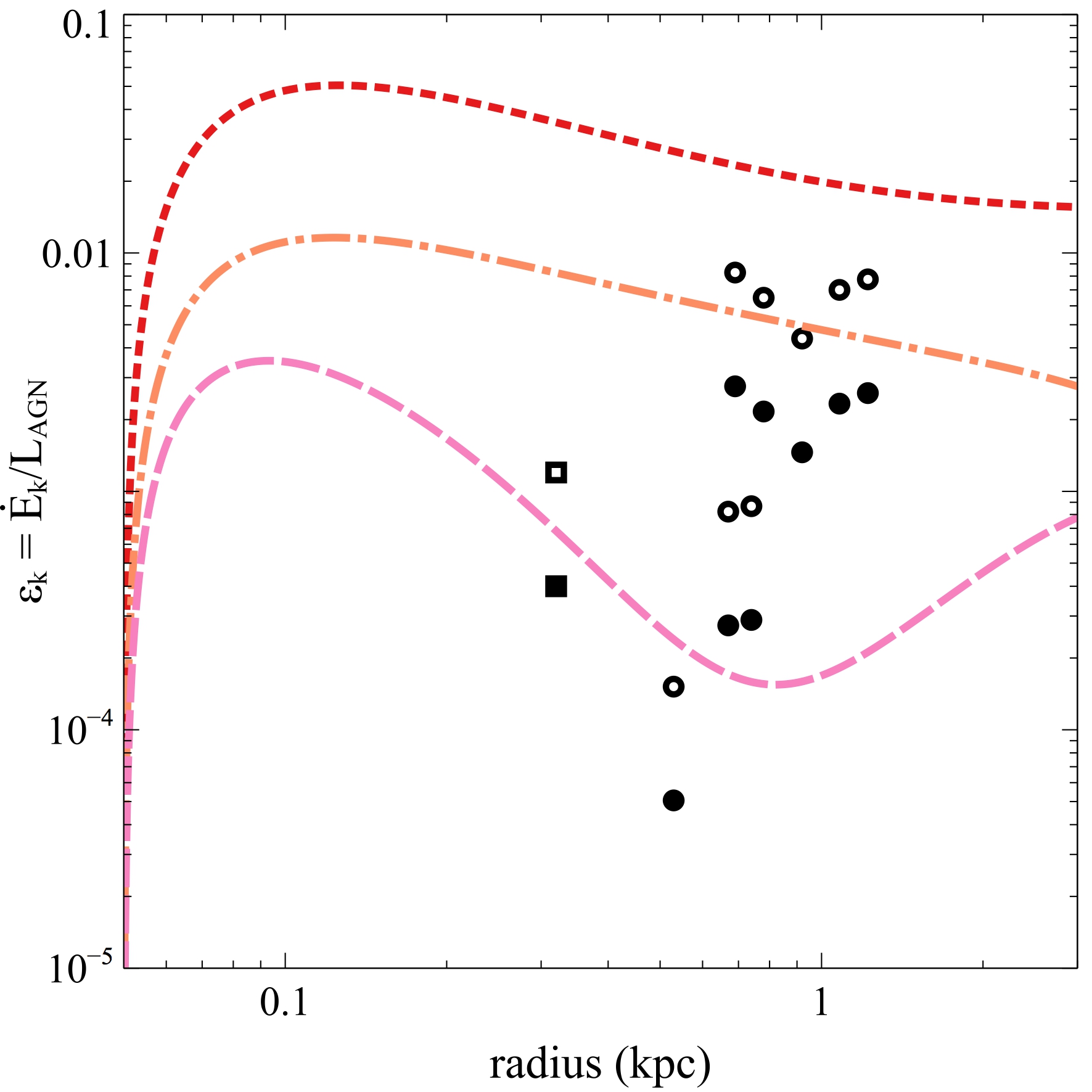}\par
    \end{multicols}
\caption{ 
Outflow energetics of dusty star-forming galaxies in the early Universe: sample of DSFG at $z > 4$ (black dots) from \citet{Spilker_et_2020} and a compact SB at $z \sim 5.7$ (black squares) from \citet{Jones_et_2019}, with molecular outflows (filled symbols) and total outflows (empty symbols). 
Left-hand panel: outflow momentum rate ($\dot{p}$) compared to the AGN-corrected (assuming a dominant AGN contribution of $\alpha_\mathrm{AGN} \sim 0.75$) SB photon momentum flux ($L_\mathrm{SF-c}/c$). 
Momentum ratio (middle panel) and energy ratio (right-hand panel) compared to AGN-driven outflows with different luminosity decay histories. Constant luminosity case: $L = 1 \times 10^{46}$ erg/s (pink dashed). Power-law luminosity decays: $L_0 = 5 \times 10^{46}$ erg/s with $t_d = 3 \times 10^5$ yr (orange dash-dot), and $L_0 = 1 \times 10^{47}$ erg/s with $t_d =  1 \times 10^6$ yr (red dotted). 
}
\label{Fig_early_DSFG}
\end{figure*}

We next consider galactic outflows in high-redshift dusty star-forming galaxies (DSFG) in the early Universe. A first survey of molecular outflows at $z > 4$ selected from a sample of gravitationally lensed DSFG is reported by \citet{Spilker_et_2020}.
A massive molecular outflow is also detected in another compact SB galaxy at $z \sim 5.7$ \citep{Jones_et_2019}. 

These DSFG are all characterised by high IR luminosities ($L_\mathrm{IR} \sim 10^{46}-10^{47}$ erg/s), with contributions from both SB and AGN components. The bolometric luminosity of the source, roughly approximated by the total IR luminosity, may be decomposed as $L_\mathrm{bol} \sim L_\mathrm{IR} = L_\mathrm{AGN} + L_\mathrm{SF}$ with $L_\mathrm{AGN} = \alpha_\mathrm{AGN} L_\mathrm{IR}$ and $L_\mathrm{SF} = (1-\alpha_\mathrm{AGN}) L_\mathrm{IR}$. 
Although the total integrated IR luminosity is often used as a proxy of the SB luminosity, the AGN contribution cannot be neglected. Actually, AGNs can contribute significantly to the far-IR emission (at a level of $\sim 40 \% - 70 \%$) in luminous quasars at high redshifts \citep{Schneider_et_2015, Duras_et_2017}. Recent observational works indicate that the AGN contribution to the IR luminosity increases with increasing $L_\mathrm{IR}$, such that galaxies can become entirely AGN-dominated at the highest $L_\mathrm{IR}$ \citep[][]{Symeonidis_Page_2018, Symeonidis_Page_2021}. As a consequence, the actual SB luminosities may be significantly lower than what derived from the total IR luminosity and should be corrected for the --possibly dominant-- AGN contribution. 

Fig. \ref{Fig_early_DSFG} (left-hand panel) shows the ratio of the outflow momentum rate ($\dot{p}$) to the AGN-corrected SB photon momentum flux ($L_\mathrm{SF-c}/c$) in high-redshift dusty star-forming galaxies. We see that most sources are located above the unity line ($\dot{p} > L_\mathrm{SF-c}/c$), and several objects also lie above the $3.5 \, L_\mathrm{SF-c}/c$ limit. A similar result is obtained when comparing the outflow kinetic power to the AGN-corrected SB luminosity output. 

As in the case of local galaxies, we examine whether the observed outflow energetics could be explained by some form of AGN luminosity evolution. In Fig. \ref{Fig_early_DSFG} (middle and right panels), we show the outflow momentum ratio and energy ratio of high-redshift DSFG compared to our AGN radiative dusty feedback models. Since the AGN contribution can be dominant in these high-luminosity sources, we consider one case with a constant AGN luminosity output and two cases with mild power-law luminosity decays (with longer characteristic decay timescales of $t_d \sim 10^5 - 10^6$ yr). The observational values can be reasonably accounted for by dusty outflows with a high degree of radiation trapping and Compton-thick initial conditions. This suggests the presence of heavily obscured and deeply buried AGN that may elude mid-IR detections \citep{Spilker_et_2020}. 

Dust-to-gas ratios of $f_\mathrm{dg} = (1 - 2) \times f_\mathrm{dg,MW}$ are adopted in the radiation pressure-driven shell models (Fig. \ref{Fig_early_DSFG}). Recent observational results indicate typical dust-to-gas ratios in the range $\sim (1/160 - 1/100)$ for dusty star-forming galaxies around the peak epoch of galaxy formation \citep{Pantoni_et_2021}. 
Large amounts of dust, with a median dust mass of $M_d \sim 10^9 M_{\odot}$, are also observed in a sample of DSFG at $1.9 < z < 6.9$ \citep{Reuter_et_2020}. 
A huge dust mass of $M_d \sim 2 \times 10^9 M_{\odot}$ is inferred in a dusty SB/obscured AGN candidate at $z \sim 6.8$ \citep{Endsley_et_2022}, while a significant dust mass of $M_d \gtrsim 10^8 M_{\odot}$ is also reported for a dust-enshrouded SB at $z \sim 7.2$ \citep{Fujimoto_et_2022}. These observations suggest a rapid dust enrichment in the early Universe, possibly due to early supernovae and grain growth \citep{Lesniewska_Michalowski_2019}. 

The momentum and energy ratios of galactic outflows in the early Universe tend to be lower than the values observed in local SB galaxies. \citet{Spilker_et_2020} argue that the modest outflow energetics of high-redshift DSFG can be fully explained by stellar feedback, without requiring any AGN contribution. 
On the other hand, they note that the high-redshift sources are not clear outliers in the plots of outflow vs. AGN properties, hence AGN cannot be ruled out as potential outflow drivers. In fact, the AGN limits are derived from mid-IR photometry, and thus the presence of buried AGN cannot be excluded. 

If the SB luminosity is overestimated (due to dominant AGN contribution), the other outflow parameters may be affected as well. 
For instance, the outflow mass loading factor, defined as the ratio of the mass outflow rate to star formation rate ($\eta = \dot{M}_\mathrm{out}/\mathrm{SFR}$) may be under-estimated. 
Unexpectedly small loading factors, in some cases well below unity ($\eta < 1$), are reported in high-redshift DSFG \citep{Spilker_et_2020, Jones_et_2019}. This may be partly due to enhanced star formation rates derived from the uncorrected IR luminosities. If the star formation rates are overestimated by a factor of a few, the actual loading factors could be correspondingly higher, bringing them more in line with theoretical expectations.


\section{Discussion}
\label{Sect_Discussion}


\subsection{AGN luminosity evolution histories}
\label{Subsect_AGN_history}

Powerful outflows observed in star-forming galaxies, without any sign of ongoing AGN activity, are usually attributed to SB-driving. 
However, it is well known that AGN variability timescales are short compared to stellar evolution timescales in galaxies. 
Galactic outflows can persist much longer than their driving phase \citep{King_et_2011, Zubovas_2018}, since AGN can vary on timescales much shorter than the flow time. The flow timescale is about $t \sim r/v \sim 10^6$ yr for outflowing gas with velocity $v \sim 1000$ km/s at a radius of $r \sim 1$ kpc. This implies that most outflows observed on kpc-scales are actually unrelated to the current nuclear activity and more likely driven by past AGN episodes. 

One of the arguments for AGN-driving rather than SB-driving is the short flow timescale measured in observational samples. 
In local star-forming galaxies, \citet{Lutz_et_2020} report flow times of $t \gtrsim$ Myr, which are difficult to explain in terms of SB activity. Typical SB events have longer characteristic timescales of $t \gtrsim 10^7$ yr, with stellar ages in the $(5-50)$ Myr range even for the most compact SBs \citep{Diamond-Stanic_et_2021}. Short $\sim$Myr-flow timescales  can be more easily accounted for by short-term AGN variability. 

Extreme luminosity variations are not uncommon in AGN host galaxies.  
Observational signatures of past AGN activity include fossil outflows and extended ionization zones, associated with otherwise quiescent nuclei. An interesting class of sources display AGN signatures on large scales ($r \gtrsim 1$ kpc), in the form of radio jets/lobes and extended emission line regions, but without any sign of ongoing nuclear activity on small scales ($r \lesssim 10$ pc). These are known as `fading AGN', whereby the central engine is already quenched (lack of nuclear X-ray and IR emissions and absence of radio core), but with large-scale ionization features or diffuse radio emission still visible on galactic scales. 

A nice example is given by Hanny's Voorwerp, where the central luminosity dropped by a factor of $\gtrsim 100$ over the last $\lesssim 10^5$ yr, leaving the extended high-ionisation cloud as a fossil record of past AGN activity \citep{Keel_et_2012}. New LOFAR data also reveal an extended radio emission from the nucleus of IC2497, suggesting that a radio jet was active in the past but has since turned off \citep{Smith_et_2022}. Arp 187 is another example of a fading or dying AGN, which likely underwent a strong luminosity decay by a factor of $\sim 10^{4}$ within $\sim 10^4$ years \citep{Ichikawa_et_2019}. 
A number of additional fading AGN candidates are reported in the local Universe \citep{Esparza-Arredondo_et_2020}.  

In some cases, the decay in central luminosity may be directly linked to the development of galactic outflows. If large amounts of gas are swept up and carried away by the outflow, the potential fuel of accreting matter is removed from the nuclear region. 
As a consequence, the accretion rate onto the central black hole falls off, and so does the radiative output. 
For instance, a ULIRG at $z \sim 0.5$ shows a strong [OIII] outflow on kpc-scales, coupled with extreme X-ray faintness and weak mid-IR emission \citep{Chen_et_2020}. 
This has been interpreted as a fading nucleus, where the primary accretion disc radiation is decaying as a result of gas removal by the powerful outflow --leading to a decline in both coronal X-ray and torus IR emissions. 

There is empirical evidence that our own Galactic Centre underwent much brighter active episodes in the past, possibly reaching Seyfert-like luminosities ($L \sim 10^{43}$ erg/s). Iron line fluorescence measurements indicate that the luminosity of Sgr A* was $L_\mathrm{X} \gtrsim 10^{39}$ erg/s in the past few hundred years \citep{Koyama_2018}. Recent 3D hydrodynamic simulations suggest that the eROSITA and Fermi bubbles, observed in X-rays and gamma-rays, may be both explained by the same Sgr A* jet event, which was active a few million years ago \citep{Yang_et_2022}. The Galactic Centre should then have been accreting at significant fractions of the Eddington rate, in striking contrast to the present-day quiescence.

In general, AGN luminosities much higher than currently observed are required to match the galactic outflow energetics. Initial AGN luminosities in the range $L \sim (10^{45} - 10^{47})$ erg/s are considered for the radiation pressure-driven shell models (Figs. \ref{Fig_local_outflows}-\ref{Fig_early_DSFG}). 
 Assuming Eddington-limited accretion, the high central luminosities imply large black hole masses. 
Several local SB galaxies have black hole masses in the range $M_\mathrm{BH} \sim (10^8 - 10^9) M_{\odot}$ \citep[][and references therein]{Zubovas_et_2022}, e.g. with IRAS 20100-4156 having an estimated mass of $M_\mathrm{BH} \sim 3.8 \times 10^9 M_{\odot}$ \citep{Harvey-Smith_et_2016}. 
Alternatively, some episodes of super-Eddington accretion could occur in some local IR-luminous galaxies \citep{Farrah_et_2022}. 

High-redshift massive DSFG --characterised by extreme star formation rates and presumably large stellar masses-- could also harbour massive black holes in their centres. Indeed, billion-solar mass black holes are already in place within the first Gyr of the Universe \citep[see the recent review by][and references therein]{Inayoshi_et_2020}. 
As an example, a black hole of mass $M_\mathrm{BH} \sim 1.6 \times 10^9 M_{\odot}$ is likely to be hosted in the heavily obscured AGN system at $z \sim 6.8$ \citep{Endsley_et_2022}. 


\subsection{An infrared view on nuclear SB and buried AGN}
\label{Subsect_IR_view}

An important element in the AGN radiative dusty feedback scenario is the trapping of reprocessed IR radiation. The outflow energetics can be significantly boosted provided that the optical depth to the reprocessed IR photons exceeds unity at the launch radius. Since the IR optical depth rapidly falls off with distance, a smaller initial radius is favoured. 
In principle, radiation trapping can occur in both nuclear SB and obscured AGN, but a geometrical difference might be expected between the two. AGN feedback originates from the point-like source at the centre, whereas stellar feedback arises from a more extended source whereby the energy deposition is more widely distributed. 
Even for the most extreme compact SB, the half-light radius is about $\sim 100$ pc \citep{Diamond-Stanic_et_2021}; while huge gas masses can be easily concentrated in the innermost few parsec-scale region around the central AGN.  
A lower limit to the initial radius could be just set by the dust sublimation radius ($r_\mathrm{sub} \gtrsim$ pc for typical AGN parameters). 

IR/sub-mm observations indicate that a major fraction of the continuum emission arises from a very small central region ($r \lesssim 10$ pc) in compact obscured nuclei, which can be optically thick even at sub-mm wavelengths. 
Similarly, the origin of the bulk IR emission seems to be confined in the nuclear region, with a median size of $r \sim (80-100)$ pc, in local ULIRGs \citep{Pereira-Santaella_et_2021}. 
If interpreted in terms of nuclear SB, extreme star formation rate densities and unusually top-heavy initial mass functions would be required \citep{Aalto_et_2019}; while local ULIRGs could be more naturally powered by buried AGN. 

On the theoretical side, radiation trapping has been extensively investigated by numerical simulations, with somewhat contrasting results reported in the literature. Early studies suggested that the reprocessed photons tend to escape through lower density channels and that Rayleigh-Taylor instabilities strongly reduce the rate of momentum transfer, such that the single scattering limit cannot be much exceeded \citep[e.g.][]{Krumholz_Thompson_2013}. However, subsequent works using refined numerical schemes showed that there can be a continuous acceleration of dusty gas, despite the development of fluid instabilities \citep{Davis_et_2014, Tsang_Milosavljevic_2015, Zhang_Davis_2017}. Simulations of radiation-driven shells indicate that the boost factor is roughly equal to the IR optical depth (except at the highest $\tau_\mathrm{IR}$), broadly confirming the simple analytic picture \citep{Costa_et_2018}. Recent RHD simulations report a good agreement between the simulated outflow propagation and the expected analytic solutions, with the inclusion of IR multi-scattering \citep{Barnes_et_2020}. 

Another concern is the relative contributions of SB and AGN to the bolometric IR luminosity of the source. As already mentioned in Section \ref{Sect_Early_Galaxies}, disentangling the two components is not trivial, and can lead to biases in the estimates of the outflow parameters. A common method used to derive the SB luminosity is based on the total integrated IR luminosity $L_\mathrm{IR} (8-1000 \, \mu m)$ measured by fitting the mid-to-far IR spectral energy distribution. But the total IR luminosity only provides an upper limit to the actual SB luminosity ($L_\mathrm{SF}$), and hence the star formation rate (SFR). This is because the AGN contribution cannot be neglected and becomes increasingly important with increasing $L_\mathrm{IR}$. Comparison between the IR luminosity functions of galaxies and AGN indicates that the two converge at the high luminosity end, suggesting that galaxies can become entirely AGN-dominated at the highest IR luminosities \citep{Symeonidis_Page_2021}. Therefore the total $L_\mathrm{IR}$ cannot be used as a reliable proxy of star formation in the most IR-luminous sources, which are more likely AGN-powered. 

It is generally assumed that the AGN contribution becomes negligible at far-IR wavelengths ($\lambda \gtrsim 100 \mu m$), and thus far-IR emission is often regarded as a reliable SFR tracer. But cold dust on galactic scales can also be heated by the central AGN itself, and such AGN-heated cold dust emission can contribute significantly to the far-IR luminosity (with the resulting far-IR colours being very similar to those of purely star-forming galaxies). Recent numerical simulations suggest that heavily enshrouded AGN can strongly boost the far-IR emission, such that the derived SFR may be overestimated by a factor of a few \citep{McKinney_et_2021}. 
So even far-IR emission cannot be considered as a safe proxy of star formation. Beyond some critical luminosity, $L_\mathrm{IR}$ will no longer trace the SFR but rather AGN activity \citep{Symeonidis_Page_2021}. At the highest IR luminosities, the SFR derived from (far)-IR emission may be systematically overestimated. 

An important corollary follows: if the IR-derived SFR are overestimated, then maximal Eddington-limited starbursts may never be reached. Exceptionally high SFR surface densities, approaching the Eddington limit for dust, seem to be required in order to launch extreme outflows \citep{Diamond-Stanic_et_2012, Heckman_Borthakur_2016, Jones_et_2019}. But the actual rate of star formation may be lower, and there could be an upper limit to the SFR that can be derived from broadband IR photometry [e.g. quantitative estimates of such maximum credible SFR at different redshifts are quoted in \citet{Symeonidis_Page_2021}]. Therefore extreme starbursts, capable of powering extreme outflows, could be much rarer than previously thought. In our picture, such extreme SB are not required, and ordinary AGN activity is sufficient to power the observed galactic outflows.


\section{Conclusion}
\label{Sect_Conclusion}

A number of works previously hinted at the possibility of relic AGN contributing to outflow-driving, but no proper constraint has been derived. Here, we try to quantify the AGN contribution in powering galactic outflows in SB galaxies (without signs of ongoing AGN activity). Overall, we find that the outflow energetics observed in many star-forming galaxies are difficult to explain in terms of pure stellar feedback. Alternatively, galactic outflows in both local SB galaxies and high-redshift DFSG could be explained by AGN feedback with different luminosity decay histories and varying degree of radiation trapping. If this is the case, then AGNs could play an even more important role than usually assumed in regulating galactic outflows across cosmic time.  


\section*{Acknowledgements }

WI acknowledges support from the University of Zurich.


\section*{Data availability}

No new data were generated or analysed in support of this research.


\bibliographystyle{mn2e}
\bibliography{biblio.bib}


\label{lastpage}

\end{document}